\documentclass [12pt]{article}
\usepackage{url}
\usepackage{hyperref}
\usepackage{amssymb}
\usepackage{amsmath}
\makeatletter
\def\section{\@startsection {section}{1}{\z@}{-3.5ex plus -1ex minus
 -.2ex}{2.3ex plus .2ex}{\large\bf}}
\def\subsection{\@startsection{subsection}{2}{\z@}{-3.25ex plus -1ex
minus -.2ex}{1.5ex plus .2ex}{\normalsize\bf}} \makeatother
\makeatletter

\@addtoreset{equation}{section}

\makeatother

\def\r{\rho}
\def\f{\phi}

\def\a{\alpha}
\def\b{\beta}
\def\d{\delta}
\def\k{\kappa}
\def\m{\mu}
\def\n{\nu}
\def\p{\pi}

\def\o{\omega}

\def\D{\Delta}
\def\G{\Gamma}

\def\be{\begin{equation}}
\def\ee{\end{equation}}
\def\bea{\begin{eqnarray}}
\def\eea{\end{eqnarray}}

\begin{document}
\begin{titlepage}
\vfill
\begin{flushright}
LU TP 04-13 \\March 2004
\end{flushright} \vspace{1cm}

\rightline{} \rightline{}
\vskip 1.8cm
\centerline{\Large \bf The Geometry of Fractional D1-branes}
 \vskip 1.4cm \centerline{\bf Daniel Bundzik${}^{a,b}$ and
Anna Tollst\'{e}n${}^a$} \vskip .8cm \centerline{$^a$ \sl School
of Technology and Society} \centerline{\sl Malm\"{o} University,
Citadellsv\"{a}gen 7, S 205 06 Malm\"{o}, Sweden}
 \centerline{\tt
Daniel.Bundzik@ts.mah.se,~~Anna.Tollsten@ts.mah.se}
 \vskip 0.4cm  \centerline{ $^b$ \sl
Department of Theoretical Physics 2} \centerline{ \sl Lund
University, S\"olvegatan 14A, S 223 62 Lund, Sweden}  \vskip 4cm
\begin{abstract}
We find explicit solutions of Type IIB string theory on
$\mathbb{R}^4 / \mathbb{Z}_2 $ corresponding
to the classical geometry of fractional
D1-branes. From the supergravity solution obtained, we
capture perturbative information about the running
of the coupling constant and the metric on the moduli space of ${\cal
N}=4$, $D=2$ Super Yang Mills.
\end{abstract}
\end{titlepage}
\renewcommand{\thefootnote}{\arabic{footnote}}
\setcounter{footnote}{0} \setcounter{page}{1}


\section{Introduction}

The success of gauge theories  to describe interactions in QED and
QCD indicate the possibility that all fundamental interactions in
Nature are of gauge type. Despite many challenging results of
non-perturbative field theories, calculations are stuck at the
perturbative level. Progress in string theory has opened up new
perspectives. As a consequence, new important perturbative and
non-perturbative results have been obtained. Supersymmetric gauge
theories can be seen as embedded in a higher dimensional string
theory containing D-branes. On the one hand, the lightest open
string excitations can be viewed as gauge fields living in the
world volume of the D-brane. On the other hand, the lightest
closed string modes correspond to D-brane solutions of
supergravity. From the duality between the open string loop
channel and the closed string tree channel one hence expects a
possible relation between gauge theory and supergravity in
general. The first exact gauge/gravity correspondence was
conjectured by Maldacena, proposing that $\mathcal{N}=4, D=4$
Super Yang-Mills theory (SYM) is dual to Type IIB string theory
compactified on $AdS_{5} \times S^{5}$\cite{Maldacena}.

To extend the AdS/CFT correspondence to non-conformal theories
with less supersymmetry, one can study string theories with wrapped
D-brane configurations in the vicinity of singularities on orbifold or
conifold backgrounds. The number of supercharges which are preserved,
and hence the possible SYM theory,
is decided by the details of the particular background. The way conformal invariance is
broken depends on how the D-branes are wrapped around the
singularity.

In order to study the wrapped D-branes alone, we should decouple
all other states. Since the mass of a static, wrapped D-brane is
proportional to the volume it encircles times the mass of the
``normal" string states, we should make this volume very small. In
the limit of vanishing volume these light, wrapped brane states
become massless and correspond to particles in the uncompactified
space-time. One only expects perturbative features of the gauge
dual from this singular geometry. When keeping the volume finite
non-perturbative effects, such as gaugino condensate and instanton
effects, occur.

A general feature of fractional D3-branes on orbifold fixed
points\cite{BDFLMP, BDFLM,BDFM} or at conical
singularities\cite{Klebanov-Nekrasov,Klebanov-Tseytlin}, is the
presence of naked singularities at small radial distance. The
fractional brane acts as a source for a twisted field which
represents the flux of an NS-NS two-form through the two-cycle.
This twisted scalar field gets radial dependence and is
interpreted, in the gauge dual, as the running coupling constant
in the IR.

In some cases, the IR singularity can be avoided by considering
wrapped D5-branes on non-vanishing Calabi-Yau
two-spheres\cite{Maldacena:2000yy,dario,zaffa,DiVecchia:2002ks},
or deformed conifolds\cite{Klebanov:2000hb}. In both these
situations, the gauge theory interpretation of chiral symmetry
breaking and gaugino condensate is controlled by a single function
in the gravitational counterpart. Moreover, it was shown in
ref.\cite{DiVecchia:2002ks} that the occurrence of
non-perturbative instanton corrections in ${\cal N}=1$ SYM smooth
out the running of the coupling constant in the IR and the theory
is thus without divergences. For a more detailed discussion and
complete reference list see for instance the review articles
\cite{Bertolini:2003iv} and \cite{Herzog:2002ih}.

In this article we will use the gauge/gravity correspondence to
study ${\cal N}=4$ SYM in $D=2$. In Section~2 we consider the
action of Type IIB string theory on $\mathbb{R}^{1,5}\times
\mathbb{R}^4 / \mathbb{Z}_2 $ using the wrapping ansatz for the
fractional D1-brane. In Section~3 we find solutions to the
equations of motion. These solutions can be expressed as a warp
factor for the untwisted fields and a radially dependent function
for the twisted fields. In Section~4 the singular fractional
D1-brane geometry is probed. Before reaching the singularity the
enhan\c{c}on radius is revealed and the breakdown of supergravity
is discussed. From the probe analysis the running Yang-Mills
coupling constant is extracted. In Section~5 we show that the
one-loop running gauge coupling for the two-dimensional gauge
theory, using the background field method, is in exact agreement
with the running coupling constant obtained from probe analysis.
The explicit equations of motion can be found in the Appendix.


\section{Action on the Orbifold}

The action of Type IIB supergravity in ten dimensions can be written (in
the Einstein frame) as\footnote{The conventions in this paper for
curved indices and forms are: $\varepsilon_{0\ldots 9}=+1$,
signature $\left(-,+^{9}\right)$, \\
$\omega_{(n)}=\frac{1}{n!}\omega_{\mu_{1}\ldots\mu_{n}}dx^{\mu_{1}}
\wedge \ldots  \wedge dx^{\mu_{n}}$ and
${^{\ast}}\omega_{(n)}=\frac{\sqrt{-detG}}{n!(10-n!)}\omega_{\mu_{1}\ldots\mu_{n}}
\varepsilon^{\mu_{1}\ldots\mu_{n}}_{\hspace{0.85cm}\nu_{1}\ldots\nu_{10-n}}dx^{\nu_{1}}
\wedge \ldots  \wedge dx^{\nu_{10-n}}$}

\bea
S_{IIB}\!\!\!& =\! \!\!&\frac{1}{2\kappa^{2}_{10}}\left\{\int
d^{10}x \sqrt{-\textnormal{det} G} R -\frac{1}{2} \int \left[
d\phi\!\wedge\! {^\ast}d\phi + e^{-\phi} H_{(3)}\!\wedge \!
{^\ast}H_{(3)} + e^{2\phi} F_{(1)}\!\wedge\! {^\ast}F_{(1)}
\right. \right. \nonumber
\\  && \hspace*{1.2cm} \left. \left. + \; e^{\phi}
\widetilde{F}_{(3)}\! \wedge \!{^\ast}\widetilde{F}_{(3)} +
\frac{1}{2}\widetilde{F}_{(5)}\! \wedge \!
{^\ast}\widetilde{F}_{(5)}- C_{(4)}\wedge H_{(3)}\wedge F_{(3)}
\right] \right\} , \eea where the field strengths \be
H_{(3)}=dB_{(2)}, \hspace{1cm} F_{(1)}=dC_{(0)}, \hspace{1cm}
F_{(3)}=dC_{(2)}, \hspace{1cm} F_{(5)}=dC_{(4)}, \ee correspond to
the NS-NS 2-form potential and the R-R 0-form, 2-form, and 4-form
with \be \widetilde{F}_{(3)}=F_{(3)}+C_{(0)}\wedge H_{(3)},
\hspace{1cm} \widetilde{F}_{(5)}=F_{(5)}+C_{(2)}\wedge H_{(3)}.
\ee The field $\widetilde{F}_{(5)}$ is self-dual, i.e
$\widetilde{F}_{(5)}= {^\ast}\widetilde{F}_{(5)}$, which condition
can only be imposed on the equations of motion. The overall factor
$\k_{10} =8\p^{7/2}g_s {\a^{\prime}}^2$ is written in terms of the
string coupling constant $g_s$ and $\a^{\prime}=l_{s}^{2}$ where
$l_s$ is the string scale.

Type IIB supergravity is now studied on the orbifold,
$\mathbb{R}^{1,5} \times \mathbb{R}^{4}/ \mathbb{Z}_{2}
\label{orbifold} $. A fractional D1-brane arises when a D3-brane
is wrapped on a compact 2-cycle of an ALE-manifold, wherupon we
take the orbifold limit \cite{DDG}. Although the cycles shrink to
zero size in the orbifold limit the fractional brane can exist
because the non-vanishing $B_{(2)}$-flux persists and is therefore
keeping the brane tensionful \cite{Merlatti-Sabella,Bertolini-Di
Vecchia-Marotta}. Since the four-form $C_{(4)}$ couples to the
D3-brane, the ``wrapping ansatz" for the fractional D1-brane is
\be B_{(2)}=b\o_2, \hspace{1cm} C_{(4)}=\widehat{C}_{(2)} \wedge
\o_2, \label{wrapping-ansatz} \ee
 where $\o_2$ is the anti-self dual
2-form related to the vanishing 2-cycle at the orbifold fixed
point. The scalar field $b$ and the 2-form $\widehat{C}_{(2)}$
will be refered to as ``twisted" fields since they correspond to
the massless states of the twisted sector of Type IIB string
theory on the orbifold.

The fractional branes are free to move in the flat transverse
directions but are forced to stay on the fixed orbifold
hyperplanes $x^{6,7,8,9}=0$, and the corresponding twisted fields are
functions of directions transverse to the orbifold,
$\r\equiv\sqrt{(x^2)^2 +\ldots +(x^5)^2}$. The bulk branes can
move freely in the orbifold directions, and the untwisted fields
are instead functions of directions transverse to the fractional
D1-brane world-volume, i.e. $r\equiv\sqrt{(x^2)^2 +\ldots
+(x^9)^2}$.

It is here appropriate to list the notation of indices used
throughout this paper. The indices for the world-volume are
denoted by $\a,\b=0,1$. The transverse space $i,j=2,\ldots,9$
consists of four flat directions $a,b=2,\ldots,5$ plus four
orbifold directions $\m,\n=6,\ldots,9$.

The fractional branes couple to closed string states. Using the
boundary state formalism\footnote{For a good review of the
boundary state formalism see for an example ref.\cite{Di
Vecchia-Liccardo}.} one can determine which fields couple to the
branes. In ref.\cite{Merlatti-Sabella} the authors study how
fractional branes in general couple to boundary states  and, in
particular, it was found that, in the the fractional D1-brane
case, the boundary state emits the NS-NS graviton $G_{ij}$ and the
dilaton $\phi$ and the R-R 2-form $C_{(2)}$  in the untwisted
sector. In the twisted sector, the two-form $\widehat{C}_{(2)}$
and the scalar $\tilde{b}$ couple to the boundary. $\tilde{b}$ is
the fluctuation part of $b$ around the background value
characteristic of the $\mathbb{Z}_2$ orbifold \cite{Aspinwall,
BCR}, $b=2\pi^2 \a^{\prime} +\tilde{b} $.

Inserting the ``wrapping ansatz"  (\ref{wrapping-ansatz}) into the
action of Type IIB string theory we obtain the action

\bea
S_{orbifold}\!\! & = \!\!
&\frac{1}{2\kappa^{2}_{orb}}\left\{\int d^{10}x
\sqrt{-\textnormal{det} G} R - \frac{1}{2}\int_{10} \left[
d\phi\wedge {^\ast}d\phi + e^{\phi} dC_{(2)}\wedge
{^\ast}dC_{(2)}\right]
 \right.  \nonumber
\\  &&   \left. -\frac{1}{2}\int_{6} \left[ e^{-\phi}d\widetilde{b}\wedge
{^{\ast_6}}d\widetilde{b}+\frac{1}{2}G_3\wedge {^{\ast_{6}}}G_3+
\widehat{C}_{(2)} \wedge d\widetilde{b}\wedge dC_{(2)} \right]
\right\}   \label{Action-orbifold} \eea on the orbifold. Here we
have introduced $\k_{orb}=\sqrt{2}\k_{10}$ and an anti-self dual
3-form defined as $G_3=d\widehat{C}_{(2)}+C_{(2)}\wedge db$. The
anti-self dual orbifold 2-cycles are normalized to \be \int \o_2
\wedge{^{\ast}} \o_2 = - \int \o_2 \wedge \o_2 =1. \ee It is
straightforward to show that the action (\ref{Action-orbifold}) is
consistent with the equations of motion of the full Type IIB
supergravity.

The boundary action is \bea S_{boundary} &=&
\frac{1}{2\kappa^{2}_{orb}}\left\{ -\frac{2T_1
\kappa_{orb}}{\sqrt{2}} \int dx^2
e^{-\phi/2}\sqrt{-\textnormal{det}G_{\a
\b}}\left(1+\frac{1}{2\pi^2 \a^{\prime}}\widetilde{b}\right)
\right. \nonumber \\ && \hspace*{1.2cm} \left. + \frac{2T_1
\kappa_{orb}}{\sqrt{2}}\int_{\mathcal{M}_2} \left[
C_{(2)}\left(1+\frac{1}{2\pi^2
\a^{\prime}}\widetilde{b}\right)+\frac{1}{2\pi^2
\a^{\prime}}\widehat{C}_{(2)}\right]\right\}
\label{Action-boundary} \eea where $T_p =
\sqrt{\pi}(2\pi\sqrt{\a^{\prime}})^{(3-p)}$ is the normalization
of the boundary state related to the brane tension in units of the
gravitational coupling constant and $\mathcal{M}_2$ is the world
volume of the fractional brane.

\section{The Ansatz and the Classical Solutions}

To find the classical solution of the low-energy string effective
action (\ref{Action-orbifold}) with boundary term
(\ref{Action-boundary}), we make the ansatz that the geometry of
the fractional D1-brane is described with the extremal metric in
the Einstein frame: \be (ds)^2=H^{-3/4}\eta_{\a \b}
dx^{\a}dx^{\b}+H^{1/4}\delta_{i j} dx^{i}dx^{j}
\label{ansatz-metric}. \ee The harmonic function $H$ is a function
of the transverse world volume directions. The ansatz for the
untwisted  2-form and the dilaton are \be
C_{(2)}=\left(\frac{1}{H}-1\right)dx^{0}\!\wedge dx^{1},
\hspace{1.5cm} e^{\phi}=H^{1/2}, \label{ansatz-untwisted} \ee
while the twisted fields $\widehat{C}_{(2)}$ and scalar field $b$
are assumed to have the form \be \widehat{C}_{(2)}=f
dx^{0}\!\wedge dx^{1} +\widehat{C}_{ab}dx^{a}\!\wedge dx^{b},
\hspace{1.5cm} b=f+{\mbox{constant}}. \label{ansatz-twisted} \ee
The function $f$ depends on the directions tranverse to the
orbifold. The axion field $C_{(0)}$ is assumed to be zero in
agreement with the ``wrapping ansatz" (\ref{wrapping-ansatz}) and
the requirement $C_{\m\n}=0$. This can be concluded from the
equation of motion for the axion field.

The above ansatz implies that the solution is  restricted to a
subspace of the complete moduli space of solutions. To relax the
self-consistent condition $\widehat{C}_{\a a}=C_{\a a}=0$ might
give a more general set of solutions. Note, however, that the
components $\widehat{C}_{ab}$ differ from zero. This is a
necessary requirement to sustain the anti-self duality of
$G_{(3)}$ which leads to the condition
\be
\partial_a \widehat{C}_{bc}+\partial_b \widehat{C}_{ca}+\partial_c
\widehat{C}_{ab}=-\varepsilon_{abc}^{\hspace{0.4cm}d}\partial_d f.
\ee Solutions to this relation can be interpreted as a solitonic
brane and look like generalized Dirac monopoles.

In the Appendix the equations of motion and more details on their
solution are presented. The equations for the twisted fields $\tilde{b}$ and
$\widehat{C}_{01}$ both give, after lengthy calculations,
\be
\partial_a \partial^a f  -K\d^{4}(x^{2,\ldots,5})= 0. \label{master-twisted}
\ee The constant is $K=T_1 \kappa_{orb}/\sqrt{2}\pi^2
\a^{\prime}$. In a similar fashion, the equations for the
untwisted fields; the metric $G_{ij}$, the dilaton $\phi$ and the
the R-R 2-form $C_{01}$ are all equivalent to \be
\partial_i \partial^i H + \partial_a f \partial^a f \d^{4}(x^{6,\ldots,9})
+\D\d^{8}(x^{2,\ldots,9})= 0, \label{master-untwisted}
\ee
where $\Delta=\sqrt{2}T_1 \kappa_{orb}$.
The singular terms of both equations are
source terms coming from the boundary action truncated to the first
order in the fluctuations around the background values of the fields.

To solve the equations (\ref{master-twisted}) and
(\ref{master-untwisted}) standard  Fourier transform techniques
are used with the resulting solutions \be
f(\r)=-\frac{K}{(2\pi)^2}\frac{1}{\r^2} \label{solution-twisted}
\ee for the twisted fields and \bea H\!\! &=&\!\!
1+\frac{\D}{2\pi^4}
\frac{1}{r^6}+\frac{K^2}{2\pi^6}\frac{1}{r^6}\left[\frac{1}{\epsilon^2}+
3\frac{r^2-2\r^2}{r^{4}}\ln \frac{(r^2-\r^2)\epsilon^2}{r^4}
\right. \nonumber  \\ && \hspace{3.7cm} \left.
+\frac{2}{r^2}-\frac{10\r^2}{r^4} +\frac{1}{2(r^2-\r^2)} \right]
\label{solution-untwisted} \eea for the untwisted fields. The
presence of the cutoff $\epsilon$ reflects the unknown
short-distance physics in directions transverse to the orbifold.
Another indication of this unknown physics is the presence of the
enhan\c{c}on radius which is discussed in the following section.

It is interesting to note that although the warp factor, $H$,
appears to differ very much from the expression in the case of
fractional D3-branes\cite{BDFLMP}, they are actually very similar.
They both contain one term which is just the spherical solution to
the Laplacian in $9-p$ dimensions with a point source, and the
remaining terms originate from the same expression in terms of a
$5-p$-dimensional integral \be
\frac{\G\left(\frac{7-p}{2}\right)\G^2\left(\frac{5-p}{2}\right)}{16\p^\frac{19-3p}{2}}
\int\frac{d^{5-p}u}{u^{8-2p}((u-x)^2+r^2-\r^2)^\frac{7-p}{2}}
\label{H-int} \ee where $(u-x)^2=\d_{ab}(u^a-x^a)(u^b-x^b)$. It
would be interesting to find out if this solution is valid for
fractional D$p\,$-branes in general.
\section{Probe analysis of the fractional brane solution}

In this section we wish to relate our result to the non-conformal
extension of the gauge/gravity-correspondence and to compare the
supergravity solution with the low-energy dynamics of the gauge
theory living on the fractional brane. The previously found
background, consisting of $N$ fractional D1-branes, is approached
by a slowly moving fractional D1-brane probe. To find the
gauge/gravity-relations we identify the gauge theory Higgs field
$\Phi^{a}$ with the transverse directions $x^{a}$ on the
supergravity side through the relation
$x^{a}=2\pi\a^{\prime}\Phi^{a}$. The probe action is defined as
the boundary action (\ref{Action-boundary}) expanded to second
order in the Higgs field. Expressed in static gauge,
$x^{0}=\xi^{0}$, $x^{1}=\xi^{1}$ and $x^{a}=\xi^{a}(x^{0})$, the
probe action becomes \be S_{probe}=  - \frac{T_1}{4\kappa_{10}}
V_2 -(2 \pi \a^{\prime })^2
\frac{T_1}{4\kappa_{10}}\left(1+\frac{\tilde{b}N}{2 \pi^2
\a^{\prime }}\right)\int d^{2}\xi
\frac{1}{2}\partial_{\a}\Phi^{a}\partial^{\a}\Phi^{b}\delta_{ab} .
\label{probe-action}\ee The first term is a constant, which shows
that all position dependent terms have cancelled. This is related
to the fact that fractional branes are BPS states and hence there
is no force between the probe and the source. The second order
term survives which enables us to define a non-trivial
four-dimensional metric on the moduli space  \be g_{ab}=(2 \pi
\a^{\prime })^2 \frac{T_1}{4\kappa}\left(1+\frac{\tilde{b}N}{2
\pi^2 \a^{\prime }}\right)\delta_{ab} =  \frac{\pi \a^{\prime
}}{2g_s }\left(1-\frac{4g_s \a^{\prime
}N}{\rho^2}\right)\delta_{ab}. \label{metric-moduli-space} \ee In
the last step we have inserted our explicit solution
(\ref{master-untwisted}). From the second term in the probe action
(\ref{probe-action}), which is interpreted as the gauge field
kinetic term, the running of the Yang-Mills coupling constant can
be extracted. It equals \be \frac{1}{g_{YM}^{2}(\rho)}=
\frac{1}{g_{YM}^{2}(\infty)}\left(1-g_{YM}^{2}(\infty)\frac{2\pi
\a^{\prime \; 2}N}{\rho^2} \right), \ee where the bare coupling
constant is defined as $g_{YM}^{2}(\infty)=2 g_{s}/\pi
\a^{\prime}$. If we now change the scale parameter to $\rho=2\pi
\a^{\prime } \mu$, we obtain the running coupling constant \be
\frac{1}{g_{YM}^{2}(\mu)}=
\frac{1}{g_{YM}^{2}(\infty)}\left(1-g_{YM}^{2}(\infty)\frac{N}{2\pi
\mu^2} \right). \ee for our fractional D1-branes. In the following
section this result will be compared to the running coupling
constant of $ \mathcal{N}=4$, $D=1+1$ super Yang-Mills theory.

To end this section we note that when the probe approaches the
radius $\rho=\rho_{e}$ where \be \rho_{e}=\sqrt{4g_{s}
\a^{\prime}N}, \ee the metric (\ref{metric-moduli-space}) on the
moduli space vanishes which means that the fractional brane
becomes tensionless. This is the enhan\c{c}on radius
\cite{Johnson-Peet-Polchinski}. For values $\rho < \rho_{e}$ the
tension becomes negative and hence undefined. The supergravity
solutions can not be trusted inside the enhan\c{c}on radius. If we
insert the value for $\rho_{e}$ into the solution
(\ref{solution-twisted}) for the $\widetilde{b}$ field, we find it
equal to the background value for the $b$ field with opposite
sign. This means that at the enhan\c{c}on radius the $b$ field
vanishes. If we express the Yang-Mills coupling constant in terms
of the $b$-field \be \frac{1}{g_{YM}^{2}(\rho)}= \frac{1}{4\pi
g_s} \int_{\Sigma_{2}}B_{(2)}= \frac{b}{4\pi g_s}, \ee we see that
at the enhan\c{c}on radius the coupling constant $g_{YM}$ goes to
infinity. To overcome this artifact one should remember that the
supergravity action is truncated to first order. This suggests
that when the probe approaches the enhan\c{c}on radius new
physical degrees of freedom, which extrapolate the reliability of
supergravity to smooth geometries, become important. The lack of a
trustworthy solution inside the enhan\c{c}on radius means that we
cannot predict the infrared behavior of the dual non-conformal
gauge theory within this framework.

\section{The running coupling constant of $ \mathcal{N}\mathbf{=4}$, $\mathbf{D=2}$ SYM }

The background field method is an efficient approach to calculate
the Yang-Mills one-loop running gauge coupling for a
$D$-dimensional field theory. The standard procedure is to write
down a  Lagrangian, gauge invariant even after gauge-fixing, with
a background external gauge field which is a solution to the
classical field equations. From the effective
action\cite{DiVecchia:2001uc}

\be S_{eff}=\frac{1}{4}\int d^D x F^2 \left(\frac{1}{g_{YM}^{2}} +
I\right), \ee the quadratic terms in the gauge fields can then be
extracted with \be I= \frac{1}{(4\pi)^{D/2}}\int_{0}^{\infty} ds
\frac{e^{-\mu^2
 s}} {s^{D/2 -1}} R.
\ee Here $\mu$ is the mass of the fields and \be R=
2\left[\frac{N_s}{12}c_s + \frac{D-26}{12}c_v +
\frac{2^{[D/2]}N_f}{6}c_f \right] . \ee The bracket means the
integer part, that is $[D/2]=D/2$ if $D$ is even and
$[D/2]=(D-1)/2$ if $D$ is odd. The constants $c$ are the
normalization of the generators of the gauge group with
$Tr(\lambda^a \lambda^b)=c\delta^{ab}$ and depend on the
representation under which the scalars, vectors and fermions
transform. $N_s$ and $N_f$ are the numbers of scalars and Dirac
fermions in the theory.

For the specific case of fractional D1-branes, there are $N_s=4$ scalars
and $N_f =2$ Dirac fermions in a $ \mathcal{N}=4$,
$D=1+1$ super Yang-Mills theory. If we choose the gauge group to
be $SU(N)$ the scalars and Dirac fermions are in the adjoint
representation which implies that $c_s = c_v = c_f = N$. With all
this in hand, we find for $D=1+1$
\be
I=
\frac{1}{4\pi}\int_{0}^{\infty} ds \; e^{-\mu^2
 s} R = \frac{1}{4\pi \mu^2}R
\ee with  $R=-2N$. This means that $I= -N/{2\pi \mu^2}$. The
running gauge coupling constant is then \be
\frac{1}{g_{YM}^{2}(\mu)}=
\frac{1}{g_{YM}^{2}(\infty)}\left(1-g_{YM}^{2}(\infty)\frac{N}{2\pi
\mu^2} \right) \ee which is in exact agreement with what we
previously found from the fractional D1-brane solution.

We can also calculate the $\beta$-function,
\be
\beta(g_{YM}(\mu)) \equiv \mu
\frac{\partial g_{YM}(\mu)}{\partial \mu} =
g_{YM}(\infty)\left(\frac{g_{YM}(\mu)}{g_{YM}(\infty)}-\left(\frac{g_{YM}(\mu)}{g_{YM}(\infty)}\right)^3\right)
\ee
which has a UV-stable fixed point at $g_{YM}(\infty)$.

\section{Discussion}

We have shown that perturbative features of $ \mathcal{N}=4$ super
Yang-Mills in two-dimensions are qualitatively inherent in the
obtained supergravity solutions for the fractional D1-brane. The
running of the coupling constant is governed by the twisted
$b\,$-field which represents the flux of the NS-NS two-form
through the compactification two-cycle. When the geometry is
studied at sub-stringy length scales, the probe becomes
tensionless before reaching the singularity. At the enhan\c{c}on
radius the $b\,$-field vanishes and supergravity is no longer a
trustworthy description. It would be interesting to study this
short-distance physics further, in context of wrapped D3-branes
where the singular orbifold is replaced by a non-vanishing
two-sphere. One expects, in a similar manner as in
ref.\cite{Maldacena:2000yy}, that identifying the spin connection
with an external gauge field would give a (4,4) SYM theory in
D=1+1 with a corresponding gravity dual free of singularities. The
running of the coupling constant is now dependent on the volume of
the two-sphere rather than the $b\,$-field. The abelian
topological twist should be performed in the UV regime but
becomes, presumably, non-abelian in the IR which smooths out the
geometry of the supergravity solutions. This enhanced gauge
symmetry have been studied for wrapped
D5-branes\cite{dario,zaffa,DiVecchia:2002ks} and it would be
interesting to see if wrapped D3-branes share the same behaviour
and account for non-perturbative results such as gaugino
condensate and instanton effects.

\vskip 1.0cm \noindent {\large {\bf Acknowledgment}} \vskip 0.2cm
\noindent We would like to thank Paolo Di Vecchia for many useful
discussions.
\vskip 1.0cm

\section{Appendix}
In this appendix more details of the calculations are presented.  the equations of motion obtained from the action
(\ref{Action-orbifold}) with boundary terms
(\ref{Action-boundary}) are presented. Inserting the ansatz
(\ref{ansatz-metric})-(\ref{ansatz-twisted}) in these equations
yield the equations (\ref{master-untwisted}) and
(\ref{master-twisted}).

The equation of motion for the field $\widehat{C}_{(2)}$ is
\be
d{^{\ast_{6}}}G_3 - db\wedge dC_{(2)} + 2K\Omega_{4}=0,
\label{c-hat}
\ee
where
\bea
G_{(3)}&=&
d\widehat{C}_{(2)}+C_{(2)}\wedge db \nonumber \\
K &=& T_1
\kappa_{orb}/(\sqrt{2}\pi^2 \a^{\prime}) \nonumber \\
 \Omega_{4} &=& \delta(x^2)\ldots\delta(x^5)dx^2 \wedge \ldots  \wedge
dx^5.
\eea
The equation of motion for the field $b$ is
\be
d(e^{-\phi} \,
{^{\ast_{6}}}db)  +G_{(3)}\wedge d{C}_{(2)}-K
V_2\Omega_{4}=0
\ee
where $V_2$ spans the world-volume of the
fractional D1-brane.

The equation of motion for the field $C_{(2)}$ is
\be
d(e^{\phi}{^{\ast_{10}}}dC_{(2)})+db\wedge
d\widehat{C}_{(2)}\wedge \widetilde{\Omega}_{4} +\D\Omega_{8}=0
\ee
where
\bea
\D &=& \sqrt{2}T_1 \kappa_{orb}\nonumber \\
\widetilde{\Omega}_{4} &=& \delta(x^6)\ldots\delta(x^9)dx^6 \wedge
\ldots  \wedge dx^9 \nonumber \\
\Omega_{8} &=& \delta(x^2)\ldots\delta(x^9)dx^2 \wedge \ldots
\wedge
 dx^9 .
\eea

The equation of motion for the dilaton $\phi$ is
\be
d{^{\ast_{10}}}d\phi - \frac{1}{2}e^{\phi}dC_{(2)}\wedge
{^{\ast_{10}}}dC_{(2)}+ \frac{1}{2}e^{-\phi}db\wedge
{^{\ast_{6}}}db\wedge \widetilde{\Omega}_{4} +\frac{1}{2}
\D V_2\Omega_{8}=0 .
\ee

To obtain the equation of motion for the metric is not quite so simple.
It is more convenient to use the language of
components instead of forms. The equation can symbolically be
expressed as $R_{MN}=L_{MN}$ where the left-hand side is the Ricci
tensor with ten-dimensional indices. There are three cases to
consider; when the indices are
$\a,\b=0,1$, $a,b=2,\ldots,5$ and $\mu,\nu=6,\ldots,9$ (remember
that $i,j=2,\ldots,9$). The result is
\be
R_{\a\b}=\frac{3}{8}H^{-2}\left(\partial_{k}\partial^{k}H
-H^{-1}\partial_{k}H\partial^{k}H \right)\eta_{\a\b},
\ee
\be
R_{ij}=-\frac{3}{8}H^{-2}\partial_{i}H\partial_{j}H
-\frac{1}{8}\left(H^{-1}\partial_{k}\partial^{k}H
-H^{-2}\partial_{k}H\partial^{k}H \right)\d_{ij} ,
\ee
and
\be
L_{\a\b} = \left(
-\frac{3}{8}H\partial_{k}C_{01}\partial^{k}C_{01}
-\frac{1}{8}H^{-2}\partial_{c}b\partial^{c}b\;\d^{4}(x)
-\frac{1}{4}G_{c01}G^{c}_{\hspace{0.25cm}01}\d^{4}(x)
-\frac{3}{8}\D\d^{8}(x) \right)\eta_{\a\b} ,
\ee
\bea
L_{ab}&=&
\frac{1}{2}\partial_{a}\f
\partial_{b}\f  \nonumber \\ &&
-\frac{1}{2}H^2\partial_{a}C_{01}\partial_{b}C_{01}+
\frac{1}{8}H^{2}\partial_{k}C_{01}\partial^{k}C_{01}\d_{ab}
\nonumber \\ &&
+\frac{1}{2}H^{-1}\partial_{a}b\;\partial_{b}b\d^{4}(x)-\frac{1}{8}H^{-1}\partial_{c}b\partial^{c}b\;\d^{4}(x)\;\d_{ab}
\nonumber \\ && -\frac{1}{2}H
G_{a01}G_{b01}\d^{4}(x)+\frac{1}{4}HG_{c01}G^{c}_{\hspace{0.25cm}01}\d^{4}(x)\;\d_{ab}
\nonumber \\ && + \frac{1}{8}\D\d^{8}(x)\d_{ab} ,
\eea
\bea
L_{\m\n}&=&
\frac{1}{2}\partial_{\m}\f
\partial_{\n}\f
-\frac{1}{2}H^2\partial_{\m}C_{01}\partial_{\n}C_{01}+
\frac{1}{8}H^{2}\partial_{k}C_{01}\partial^{k}C_{01}\d_{\m\n}
\nonumber \\ &&
+\frac{1}{8}H^{-1}\partial_{c}b\partial^{c}b\;\d^{4}(x)\;\d_{\m\n}
+ \frac{1}{8}\D\d^{8}(x)\d_{\m\n} .
\eea
Combining these relations in an
appropriate manner gives the same equation
(\ref{master-untwisted}) just like the equations for $C_{(2)}$ and
$\phi$ do.

\end{document}